# Unified Flow and Thermal Law-of-the-wall for Type-A and B Turbulent Boundary Layers

Hongyi Xu[1], Duo Wang[1], Heng Li[1] and Bochao Cao[1]

[1]*Aeronautics and Astronautics Department,
Fudan University, Shanghai, China, 200433
E-mail address: Hongyi_Xu@fudan.edu.cn*

Systematic study of the existing statistical data from direct numerical simulations (DNS) lead to a logical and important classification of generic turbulent boundary layers (TBL), namely Type-A, -B and -C TBL, based on the distribution patterns of time-averaged local wall-shear stress. Among these three types, Type-A and B TBLs are investigated in terms of its universal statistical laws of velocity and temperature with independency on both Reynolds (Re) and Prandtl (Pr) numbers. With reference to the analysis by von Karman in developing the conventional law-of-the-wall for Type-A TBL, the current study first physically distinguishes the time-averaged local velocity scale from the ensemble-averaged velocity scale, and properly defines the TBL temperature scale based on wall heat-flux, similar to the well-known frictional velocity scale based on wall shear-stress. The unified law-of-the-wall formulations are then derived for the inner layer of Type-A and B thermal TBLs using these scales. The formulations are constructed by introducing the general damping and enhancing functions in inner layer, and further, these functions are extended to logarithmic coordinate based on the multiscale characteristics of TBL velocity profiles. The validations of the law are conducted both by the DNS-guided near-wall integration of Type-B TBL governing equations, which uncovered the law's *Re*- and *Pr*-independency mechanisms. The research advances the current understandings of the conventional TBL theory and develop the complete analytical forms of law-of-the-wall for Type A and B TBLs, including inner-layer, transition layer, logarithmic layer and wake layer.

Over the past century, searching for universal statistical law in turbulence has always been a persistent effort for fluid mechanics community. The von Karman law-of-the-wall [1] is well-known as one of the milestones in the theory of Prandtl's turbulent boundary layer (TBL) [2]. The law was found through an in-depth observation of the turbulence on a flat wall and a unique analysis using the time-averaged local frictional velocity as scale [1]. So far, it has been studied and proofed by many experiments [3], theoretical methods [4], as well as numerous modern Direct Numerical Simulations (DNS) for simple geometry TBLs [5, 6, 7], such as TBLs on a flat-plate, channel or circular pipe.

As the DNS studies were successfully applied to more complex geometry TBLs, the bigger database permitted a thorough interrogation of the connotation behind TBL concept. Within the context, Xu [8] first applied the methodology in [1] to the statistics of turbulence in more sophisticated TBLs, such as square and rectangular annular ducts [9], and pointed out the necessity to classify generic TBLs into three types, namely Type-A, -B and -C TBL. The classification was logically based on the distribution patterns of the time-averaged local frictional velocity [1], i.e. the TBL velocity scale in the conventional law, which opened the door to re-understand the physical essence of the law in inner layer, i.e. $U^+ = y^+$, and provoked the thinking to search for more appropriate velocity scale in Type-B TBL. By using the ensemble-averaged frictional velocity as scale, Xu [8] successfully derived the law's formulation, i.e. the mathematical form of $U^+(y^+)$, for Type-B TBL and preliminarily provided its validity using the DNS-guided near-wall integration of Type-B TBL governing equation.

Inspired by the von Karman's work [1], the current letter first derives the proper TBL scales for both fluid and thermal fields from reasoning the definitions of wall shear stress and wall heat flux. These scales are then applied to Type-B TBL governing equations, which paved the way to search for the near-wall statistical regularities of velocity and temperature with Reynolds number (*Re*) and Prandtl number (*Pr*) independency. As guided by the reliable Type-B TBL DNS data at both high and low *Re*, the study demonstrates the evident *Re* and *Pr* independency for the formulation in [8] and further reveals its formation mechanisms by quantitatively interrogating the contributions of each term in Type-B TBL governing equations. The analysis further discovers that the velocity is actually an approximate *Re* independency with the accuracy proportional to *Re* reciprocal, whereas the temperature is strictly *Re* and *Pr* independent. These findings advance the current knowledges of TBL and wall-bounded thermal turbulence.

1. Definition of Type-A, -B and -C TBL

Since TBL scaling is a critically important issue in developing the conventional law-of-the-wall [1], it is logical to classify generic TBLs in terms of the distribution patterns of time-averaged wall-shear stress ($\tau_w$), a physical quantity closely associated with TBL velocity scale. Cao & Xu [8] pointed out the importance of classification in re-understanding the conventional law-of-the-wall. The definition reads: Type-A with $\tau_w = \tau_w(x)$, or $\tau_w =$ constant; Type-B with $\tau_w = \tau_w(z)$ or $\tau_w = \tau_w(y)$ and Type-C with $\tau_w = \tau_w(x,z)$ or $\tau_w = \tau_w(x,y)$ where $x$ is the streamwise direction, $y$ or $z$ is the wall normal or spanwise direction, respectively, and $\tau_w$ is the time-averaged wall-shear stress. With the TBL classification, the conventional law was suited and well-known validated for Type-A TBL. The law's applicability to Type-B TBL were preliminarily investigated in Cao & Xu [8]. The current letter provides detailed proof and insight on the formation mechanisms of the law in the inner-layer of Type-B TBL.

## 2. Flow and Thermal TBL Scales

From the von Karman's work [1], TBL velocity and length scales were identified by nondimensionalizing the definition of wall-shear stress ($\tau_w$). The key procedures include: (1) making the left hand side of the definition be equal to unity; and then (2) reorganizing the right hand side by making the nominator and denominator in partial derivative be nondimensionalized so that the appropriate velocity and length scales are derived. The same logics can be applied to the definition of wall-heat flux, and hence the temperature and corresponding length scales for thermal TBL are obtained.

$$\tau_w = \mu \frac{\partial u}{\partial y} \rightarrow 1 = \mu \frac{\partial \frac{u}{\tau_w}}{\partial y} = \frac{\partial \frac{u}{u_\tau}}{\partial \frac{\rho u_\tau y}{\mu}} = \frac{\partial u^+}{\partial y^+} \quad (1); \quad -q_w = k \frac{\partial T}{\partial y} \rightarrow -1 = k \frac{\partial \frac{T}{q_w}}{\partial y} = \frac{\partial \frac{T}{c\rho u_\tau}}{\partial \frac{c\rho u_\tau y}{k}} = \frac{\partial T^+}{\partial y^*} \quad (2)$$

Within the context, the TBL velocity and length scales are well-known defined by: $u_\tau = \sqrt{\tau_w/\rho}$, $y^+ = \rho u_\tau y/\mu$, $u^+ = u/u_\tau$ and the thermal TBL temperature and length scales can logically be obtained as: $T_\tau = q_w/(c\rho u_\tau)$, $y^* = \rho u_\tau y/\mu \cdot c\mu/k = y^+/Pr$, $T^+ = T/T_\tau$, where $c$ is specific heat, $\rho$ is density, $\mu$ is dynamic viscosity and $Pr$ is Prandtl number. By using these appropriately defined TBL scales, including velocity, temperature and length, the scale analysis of governing equations for Type-B TBL is conducted to prove the existence of solution possessing the property of $Re$ and $Pr$ independency.

## 3. Scale analyses of Type-A and B TBL governing equations
### 3.1 Type-B TBL governing equations

Cao & Xu [8] raised an important issue to distinguish the local time-average wall shear stress ($\tau_w$) in the conventional law from the ensemble-averaged wall shear stress ($\overline{\tau_w}$) in developing the new law for Type-B TBL. Therefore, based on the definition of Type-B TBL and the newly-introduced ensemble-averaged wall-heat flux ($\overline{q_w}$), the Type-B TBL velocity and temperature scales can then be defined as $\overline{u_\tau} = \sqrt{\overline{\tau_w}/\rho}$ and $\overline{T_\tau} = \overline{q_w}/(c\rho \overline{u_\tau})$, respectively. It is worthy to note that since the ensemble-averaged quantities are no longer a function of independent variables of $(y^+, z^+)$ or $(y^*, z^*)$, and therefore the nondimensional form of time-averaged governing equations for Type-B TBL can generally be written as:

$$\frac{\partial \overline{V^+ U^+}}{\partial y^+} + \frac{\partial \overline{W^+ U^+}}{\partial z^+} = \frac{-1}{Re_\tau} \frac{\partial \bar{p}^+}{\partial x} + \left(\frac{\partial^2 \overline{U^+}}{\partial y^{+2}} + \frac{\partial^2 \overline{U^+}}{\partial z^{+2}}\right) + \left(\frac{\partial \overline{V^{+\prime} U^{+\prime}}}{\partial y^+} + \frac{\partial \overline{W^{+\prime} U^{+\prime}}}{\partial z^+}\right) \ldots Eq.3$$

$$\frac{\partial \overline{V^+ V^+}}{\partial y^+} + \frac{\partial \overline{W^+ V^+}}{\partial z^+} = -\frac{\partial \bar{p}^+}{\partial y^+} + \left(\frac{\partial^2 \overline{V^+}}{\partial y^{+2}} + \frac{\partial^2 \overline{V^+}}{\partial z^{+2}}\right) + \left(\frac{\partial \overline{V^{+\prime} V^{+\prime}}}{\partial y^+} + \frac{\partial \overline{W^{+\prime} V^{+\prime}}}{\partial z^+}\right) \ldots Eq.4$$

$$\frac{\partial \overline{V^+ W^+}}{\partial y^+} + \frac{\partial \overline{W^+ W^+}}{\partial z^+} = -\frac{\partial \bar{p}^+}{\partial z^+} + \left(\frac{\partial^2 \overline{W^+}}{\partial y^{+2}} + \frac{\partial^2 \overline{W^+}}{\partial z^{+2}}\right) + \left(\frac{\partial \overline{V^{+\prime} W^{+\prime}}}{\partial y^+} + \frac{\partial \overline{W^{+\prime} W^{+\prime}}}{\partial z^+}\right) \ldots Eq.5$$

$$\frac{\partial \overline{V^+ T^+}}{\partial y^*} + \frac{\partial \overline{W^+ T^+}}{\partial z^*} = \left(\frac{\partial^2 \overline{T^+}}{\partial y^{*2}} + \frac{\partial^2 \overline{T^+}}{\partial z^{*2}}\right) + \left(\frac{\partial \overline{V^{+\prime} T^{+\prime}}}{\partial y^*} + \frac{\partial \overline{W^{+\prime} T^{+\prime}}}{\partial z^*}\right) \ldots Eq.6$$

where $\overline{U^+} = u/\overline{u_\tau}$, $\overline{T^+} = T/\overline{T_\tau}$, $y^+ = \rho \overline{u_\tau} y/\mu$, $y^* = \rho \overline{u_\tau} y/\mu /Pr$, $z^+ = \rho \overline{u_\tau} z/\mu$, $z^* = \rho \overline{u_\tau} z/\mu /Pr$.

Since the time-averaged pressure gradient of $\partial \bar{p}^+/\partial x$ is a constant in Type-B TBL, see Cao & Xu[8], the solution from Eq. 3, 4, 5, i.e. $(\overline{U^+}, \overline{V^+}, \overline{W^+}) = (\overline{U^+}, \overline{V^+}, \overline{W^+})(y^+, z^+, 1/Re_\tau)$, is an approximate $Re_\tau$-independency as long as $Re_\tau$ is sufficiently large, and the proofs are given in Section 5 by the validation using DNS data. Needless to say, the solution from Eq. 4, i.e. $\overline{T^+} = \overline{T^+}(y^*, z^*)$, is strictly $Re_\tau$- and $Pr$-independent. These equations are the foundation to understand, prove and validate the unified inner-layer law for Type-B TBL.

### 3.2 Type-A TBL governing equations

As represented by the turbulent flow along zero-pressure gradient semi-infinite flat-plate, the time-averaged governing equation for a Type-A TBL in dimensional for can be written as :

$$\frac{\partial \overline{u}\overline{u}}{\partial x} + \frac{\partial \overline{v}\overline{u}}{\partial y} = \nu\left(\frac{\partial^2 \overline{u}}{\partial x^2} + \frac{\partial^2 \overline{u}}{\partial y^2}\right) - \left(\frac{\partial \overline{u'u'}}{\partial x} + \frac{\partial \overline{u'v'}}{\partial y}\right) \cdots Eq.7$$

$$\frac{\partial \overline{u}\overline{v}}{\partial x} + \frac{\partial \overline{v}\overline{v}}{\partial y} = -\frac{\partial \overline{p}}{\partial y} + \nu\left(\frac{\partial^2 \overline{v}}{\partial x^2} + \frac{\partial^2 \overline{v}}{\partial y^2}\right) - \left(\frac{\partial \overline{u'v'}}{\partial x} + \frac{\partial \overline{v'v'}}{\partial y}\right) \cdots Eq.8$$

$$\frac{\partial \overline{u}\overline{T}}{\partial x} + \frac{\partial \overline{v}\overline{T}}{\partial y} = k\left(\frac{\partial^2 \overline{T}}{\partial x^2} + \frac{\partial^2 \overline{T}}{\partial y^2}\right) - \left(\frac{\partial \overline{u'T'}}{\partial x} + \frac{\partial \overline{v'T'}}{\partial y}\right) \cdots Eq.9$$

Similar to the scale selection in Type-B TBL, here by defining $\overline{u_\tau} = \sqrt{\overline{\tau_w}/\rho}$ and $\overline{T_\tau} = \overline{q_w}/(c\rho\overline{u_\tau})$, therefore $\overline{U^+} = u/\overline{u_\tau}$, $\overline{T^+} = T/\overline{T_\tau}$, $x^+ = \rho\overline{u_\tau}x/\mu$, $x^* = \rho\overline{u_\tau}x/\mu/Pr$, $y^+ = \rho\overline{u_\tau}y/\mu$, $y^* = \rho\overline{u_\tau}y/\mu/Pr$ , the non-dimensional governing equation for Type-A TBL can be written as:

$$\frac{\partial \overline{u^+}\,\overline{u^+}}{\partial x^+} + \frac{\partial \overline{v^+}\,\overline{u^+}}{\partial y^+} = \left(\frac{\partial^2 \overline{u^+}}{\partial x^{+2}} + \frac{\partial^2 \overline{u^+}}{\partial y^{+2}}\right) - \left(\frac{\partial \overline{u^{+'}u^{+'}}}{\partial x^+} + \frac{\partial \overline{u^{+'}v^{+'}}}{\partial y^+}\right) \cdots Eq.10$$

$$\overline{u^+}\frac{\partial \overline{v^+}}{\partial x^+} + \overline{v^+}\frac{\partial \overline{v^+}}{\partial y^+} = -\frac{\partial \overline{p^+}}{\partial y^+} + \left(\frac{\partial^2 \overline{v^+}}{\partial x^{+2}} + \frac{\partial^2 \overline{v^+}}{\partial y^{+2}}\right) - \left(\frac{\partial \overline{u^{+'}v^{+'}}}{\partial x^+} + \frac{\partial \overline{v^{+'}v^{+'}}}{\partial y^+}\right) \cdots Eq.11$$

$$\overline{u^+}\frac{\partial \overline{T^+}}{\partial x^*} + \overline{v^+}\frac{\partial \overline{T^+}}{\partial y^*} = \left(\frac{\partial^2 \overline{T^+}}{\partial x^{*2}} + \frac{\partial^2 \overline{T^+}}{\partial y^{*2}}\right) - \left(\frac{\partial \overline{u^{+'}T^{+'}}}{\partial x^*} + \frac{\partial \overline{v^{+'}T^{+'}}}{\partial y^*}\right) \cdots Eq.12$$

It should be noted that after scaled by the selected scaling quantities, the governing equations no longer contain any flow and thermal criterion numbers such as Reynolds or Prantdl numbers, which guarrantee the analysis based on the scales using in these equations are independent on the variations of the flow and thermal criterion numbers, or universally applicable to the thermal turbulence with different criterion numbers. This is main reason to introduce the time-space averaged frictional velocity and the wall-heat flux and these governing equations then become the theoretical foundation to explore the universally applicable law's formulations. On the other hand, the time-averaged frictional velocity [$\tau_w = \tau_w(x)$] used by von Karman was obviously not suitable to be used as scale to obtained these criterion-number indenpendent governing equations.

4. Unified flow and thermal Law-of-the-wall for Type-A and B TBLs

Cao & Xu [8] made an important breakthrough by in-depth observation of mean streamwise velocity in the inner layer of Type-B TBL, which led to deriving the <u>general damping function</u> [$f(d^+) = 1 - \Delta e^{-d^+/D}$], an extension form of the van Driest damping function [9] by introducing the incremental gradient parameter of $\Delta$, and the <u>general enhancing function</u> [$f(d^+) = 1 + \Delta e^{-d^+/D}$], a logically speculated function with an enhancing nature opposite to the damping function. The von Karman inner-layer law, i.e. $\overline{U^+} = d^+$ with $d^+$ being the wall normal distance, was found capable of well representing the inner-layer streamwise velocity of Type-B TBL when corrected by the damping and enhancing functions [8]. Therefore, by using the identified TBL scales in *Eqs.* 3-6, i.e. $\overline{u_\tau}$ and $\overline{T_\tau}$ , the unified flow and thermal law-of-the-wall in TBL inner-layer can then be derived and written as:

**For flow-TBL Type-B:**

(1) Inner layer $0.0 \leqslant y^+ \leqslant \approx 10.0$

when $\tau_w/\overline{\tau_w} \leqslant 1.0$, $\overline{U^+}(y^+) = y^+[1 - \Delta(z^+)e^{-\frac{y^+}{D(z^+)}}]$ ... $Eq.7$, where $\Delta = 1 - \tau_w/\overline{\tau_w}$;

when $\tau_w/\overline{\tau_w} \geqslant 1.0$, $\overline{U^+}(y^+) = y^+[1 + \Delta(z^+)e^{-\frac{y^+}{D(z^+)}}]$ ... $Eq.8$, where $\Delta = \tau_w/\overline{\tau_w} - 1$;

(2) Transition layer $10.0 \leqslant \approx y^+ \leqslant 50.0$, δ is flow TBL thickness

(3) Logarithmic layer $50.0 \leqslant y^* \leqslant 0.15\delta$, $\delta$ is flow TBL thickness

(4) Wake layer $0.15\delta \leqslant y^* \leqslant 0.90\delta$, $\delta$ is flow TBL thickness

$y^+$ is the independent variable in the direction normal to wall and $z^+$ is in the corresponding spanwise direction;

**For thermal TBL Type-B:**

(1) Inner layer $0.0 \leqslant y^* \leqslant\approx 10.0$
when $q_w/\overline{q_w} \leqslant 1.0$, $\overline{T^+}(y^*) = y^*[1 - \Delta(z^*)e^{-\frac{y^*}{D(z^*)}}] \ldots Eq.9$, where $0 \leqslant \Delta \leqslant 1$ and $1 - \Delta = q_w/\overline{q_w}$;
when $q_w/\overline{q_w} \geqslant 1.0$, $\overline{T^+}(y^*) = y^*[1 + \Delta(z^*)e^{-\frac{y^*}{D(z^*)}}] \ldots Eq.10$ where $0 \leqslant \Delta \leqslant 1$ and $1 + \Delta = q_w/\overline{q_w}$;

(2) Transition layer $10.0 \leqslant\approx y^* \leqslant 50.0$, $\delta$ is temperature TBL thickness

(3) Logarithmic layer $50.0 \leqslant y^* \leqslant 0.15\delta$, $\delta$ is temperature TBL thickness

(4) Wake layer $0.15\delta \leqslant y^* \leqslant 0.90\delta$, $\delta$ is temperature TBL thickness

$y^*$ is the independent variable in the direction normal to wall and $z^*$ is in the corresponding spanwise direction.

It is worth to note that the parameters of $\Delta$ and D are independent on the criterion number of $Re_\tau$ and $Pr$ if *Eqs* 7-10 are the solutions of *Eqs*. 3-6, and therefore in general, $\Delta(z^+)$ and $D(z^+)$ are universally applicable parameters for any $Re_\tau$ and $Pr$.

**For flow TBL Type-A:**

(1) Inner layer $0.0 \leqslant y^+ \leqslant\approx 10.0$
when $\tau_w/\overline{\tau_w} \leqslant 1.0$, $\overline{U^+}(y^+) = y^+[1 - \Delta(x^+)e^{-\frac{y^+}{D(x^+)}}] \ldots Eq.7$, where $\Delta = 1 - \tau_w/\overline{\tau_w}$;
when $\tau_w/\overline{\tau_w} \geqslant 1.0$, $\overline{U^+}(y^+) = y^+[1 + \Delta(x^+)e^{-\frac{y^+}{D(x^+)}}] \ldots Eq.8$, where $\Delta = \tau_w/\overline{\tau_w} - 1$;

Proof of the conventional inner-layer law (viscous linear layer law) and the newly-developed inner-layer law

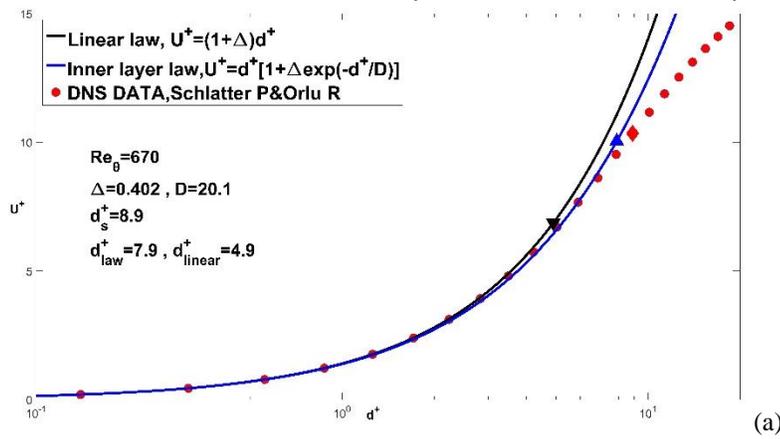
(a)

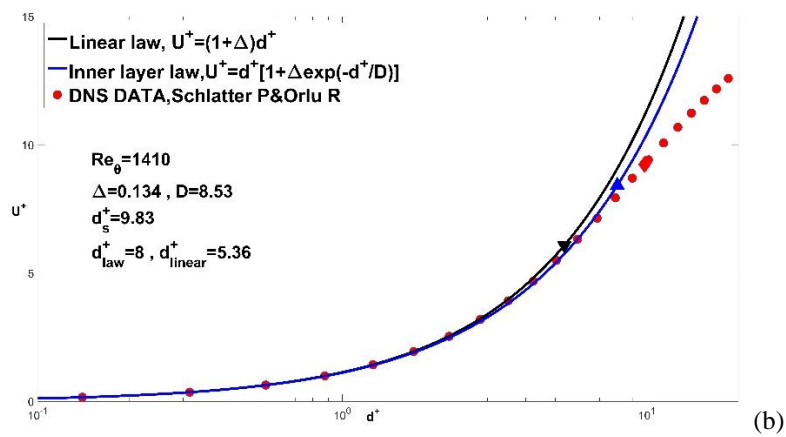
(b)

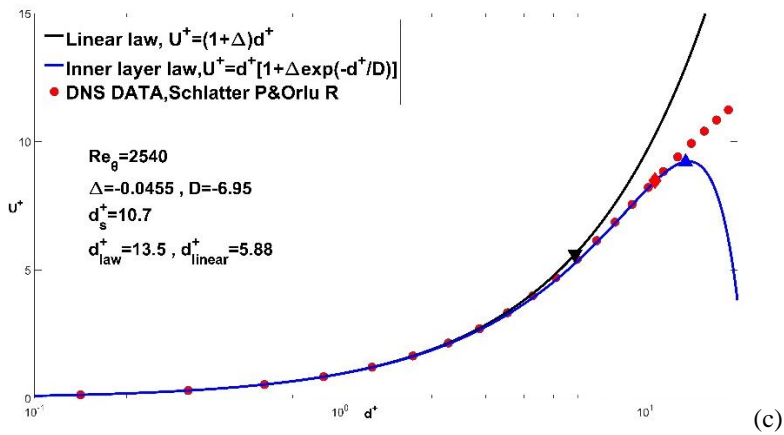
(c)

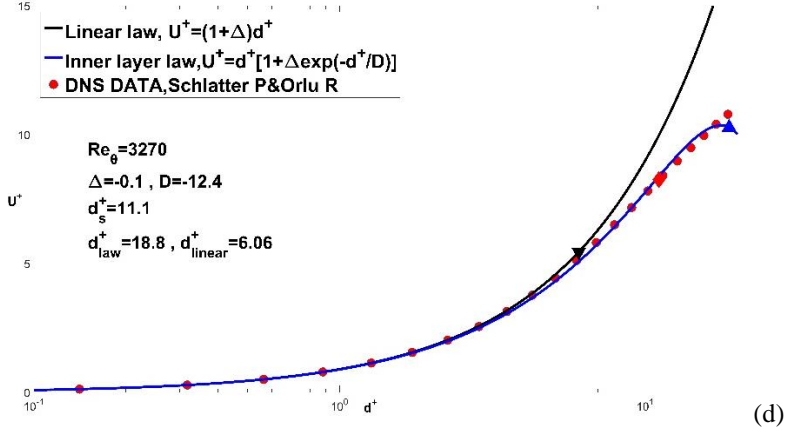

(d)

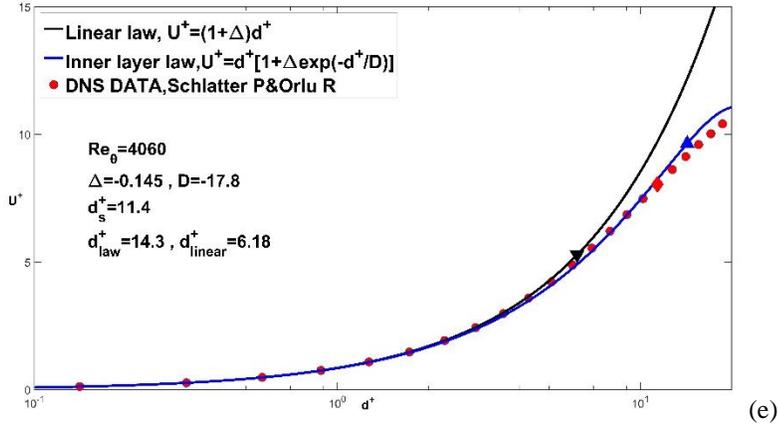

(e)

(2) Transition layer $10.0 \leqslant \approx y^+ \leqslant 50.0$
$$U^+ = U_s - \{1 + \Delta_e e^{[-\ln(y^+/y_s)/D_e]}\}[-\ln(y^+/y_s)]$$
where

(3) Logarithmic layer $50.0 \leqslant y^+ \leqslant 0.15\delta$, $\delta$ is flow TBL thickness
$$U^+ = \frac{\sqrt{1+\Delta}}{\kappa^*}\ln y^+ + \sqrt{1+\Delta}(\frac{\ln\sqrt{1+\Delta}}{\kappa^*} + C^*)$$
where

(4) Wake layer $0.15\delta \leqslant y^+ \leqslant 0.90\delta$, $\delta$ is flow TBL thickness
$$U^+ = \{1+\Delta_w e^{[\ln(y^+/y_w^+)/D_w]}\}[\ln(y^+/y_w^+)] + U_w^+$$
where

$y^+$ is the independent variable in the direction normal to wall and $z^+$ is in the corresponding spanwise direction;

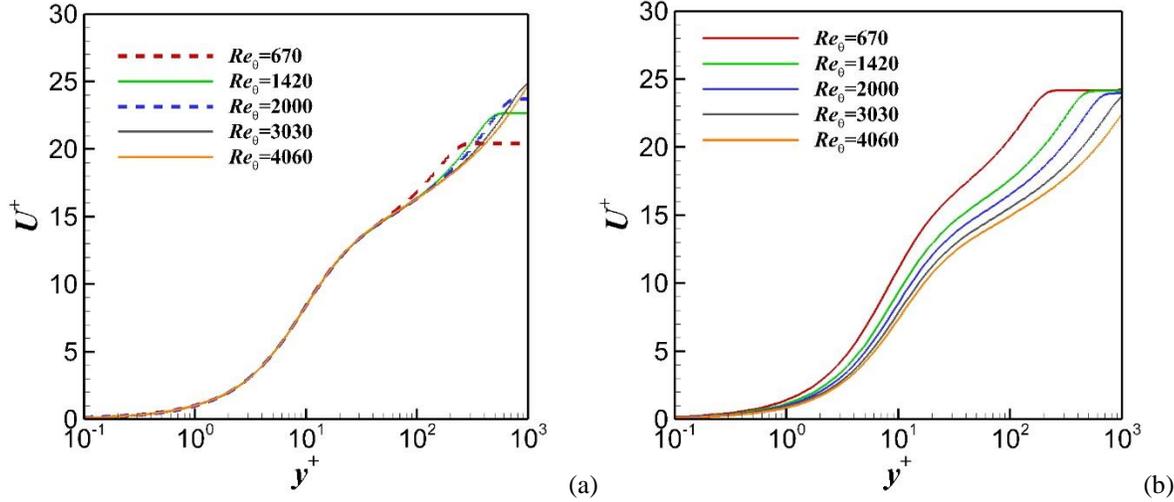

Proof of (a) the conventional Law-of-the-wall and (b) the newly-developed Law-of-the-wall using Type-A DNS data as published by Schlatter & Örlü[9].

**For thermal TBL Type-A:**

(1) Inner layer $0.0 \leqslant y^* \leqslant \approx 10.0$

when $q_w/\overline{q_w} \leqslant 1.0$, $\overline{T^+}(y^*) = y^*[1 - \Delta(z^*)e^{-\frac{y^*}{D(z^*)}}]$ ... $Eq.\,9$, where $0 \leqslant \Delta \leqslant 1$ and $1 - \Delta = q_w/\overline{q_w}$;

when $q_w/\overline{q_w} \geqslant 1.0$, $\overline{T^+}(y^*) = y^*[1 + \Delta(z^*)e^{-\frac{y^*}{D(z^*)}}]$ ... $Eq.\,10$ where $0 \leqslant \Delta \leqslant 1$ and $1 + \Delta = q_w/\overline{q_w}$;

(2) Transition layer $10.0 \leqslant \approx y^* \leqslant 50.0$

(3) Logarithmic layer $50.0 \leqslant y^* \leqslant 0.15\delta$, $\delta$ is temperature TBL thickness

(4) Wake layer $0.15\delta \leqslant y^+ \leqslant 0.90\delta$, $\delta$ is temperature TBL thickness

$y^*$ is the independent variable in the direction normal to wall and $z^*$ is in the corresponding spanwise direction.

5. Proof and Validation of the law using DNS statistical data

To prove and validate the correctness and accuracy of *Eqs.* 7-10, the DNS statistics of Type-B TBLs, such as the fully-developed turbulence in square and rectangular annular ducts (SAD and RAD) [8], are analyzed based on *Eqs.* 3-6, particularly with regard to the $Re_\tau$- and *Pr*-independency. The F- and T-TBLs with the representative $\Delta$ and D listed in Table-1 are selected to demonstrate the fitness of *Eq.* 7 and 8 to the DNS data. To profoundly understand the formation mechanisms of the law, the TBL profiles are plotted in Figs. 1-10 and are analyzed by *Eq.* 3, 6 and the relevant Reynolds stress statistics from DNS data.

6. Conclusions and discussions